\documentclass[graybox]{svmult}

\usepackage{mathptmx}       
\usepackage{helvet}         
\usepackage{courier}        
\usepackage{type1cm}        
\usepackage{makeidx}         
\usepackage{graphicx}        
\usepackage{grffile}
\usepackage{epstopdf}        
\usepackage{amsmath}
\usepackage{multicol}        
\usepackage[bottom]{footmisc}

\makeindex             

\begin{document}

\title*{OpenMP parallelization of multiple precision Taylor series method}
\author{S. Dimova$^1$, I. Hristov$^{1,a}$, R. Hristova$^1$, I. Puzynin$^2$, T. Puzynina$^2$, Z. Sharipov$^{2,b}$,\\ N. Shegunov$^1$, Z. Tukhliev$^2$\\
\vspace{1cm}
$^1$ Sofia University, Faculty of Mathematics and Informatics, Bulgaria\\
$^2$ JINR, Laboratory of Information Technologies, Dubna, Russia\\
\vspace{1cm}
Corresponding e-mails:  \hspace{0.15cm}$^a$  ivanh@fmi.uni-sofia.bg,\hspace{0.15cm} $^b$ zarif@jinr.ru}
\authorrunning{S. Dimova, I. Hristov, et al.}

%
%
\maketitle

\abstract{OpenMP parallelization of multiple precision Taylor series method is proposed.
A very good parallel performance scalability and parallel efficiency inside one computation node of a CPU-cluster is observed.
We explain the details of the parallelization on the classical example of the Lorentz equations.
The same approach can be applied straightforwardly to a large class of chaotic dynamical systems.}

\section{Introduction}
\label{sec:2}

Achieving a mathematically reliable long-term solution of a chaotic dynamical system is a difficult task due to the sensitive dependence on the initial conditions. No doubt, having a numerical procedure for achieving such solutions is of a big importance, because it  gives us a powerful tool for theoretical investigations. A breakthrough in this direction can be found in the paper \cite{Liao1} of Shijun Liao. The author of the paper considers a new numerical procedure called "clean numerical simulation", based on  the multiple precision Taylor series method \cite{Jorba}, \cite{Barrio}.

The numerical procedure  in \cite{Liao1}  works as follows. A new concept, namely the critical predictable time $T_c$ is introduced.
Practical estimations of the needed accurate decimal places $K$ and the needed  order $N$ of the Taylor method are obtained by constructing $T_c - K$ and $T_c - N$ diagrams. Using the estimated $K$ and $N$ for a given time interval, the solution is calculated. The solution obtained  is additionally verified by a new calculation with larger $K$ and $N$ over the same interval. If the two solutions coincide over the whole interval, the solution is considered to be a mathematically reliable one.

To apply the "clean numerical simulation", we have to be precise in many directions. First, we have to use a multiple precision library. In order to be effective, we need a method of highest  order of accuracy, such as Taylor series method. Low order accuracy methods require very small steps to be used and as a result we can not compute the solution in a foreseeable time.  In addition,  if we want a solution in the case of extremely large intervals, we need a
serious computational resource and a parallelization of the algorithm.

First parallelization of Liao's method is reported in \cite{par1} and later improved in \cite{par2}. A mathematically reliable solution of the Lorenz system using 1200 CPU cores, obtained in about 9 days and 5 hours, on a time interval with a record length, namely [0,10000], is given in \cite{Liao2}. It is explained in \cite{par1},\cite{par2}  that a parallel reduction of the sums, which appear when we calculate the Taylor coefficients, have to be done. This is of course the crucial observation. However, no details of the parallel version of the algorithm is given. Most likely the authors used pure MPI (Message Passing Interface) which is a distributed memory programming model.

Our goal is not to compare to the impressive simulation in \cite{Liao2}, which uses pretty large computational resource. Our goal is to present in more details a simple and effective OpenMP parallelization of the multiple precision Taylor series method, which uses a moderate computational resource, namely one CPU-node. For benchmarks, we use the results in \cite{par2}. Our model problem is the classical Lorenz system. However, the proposed approach is rather general, and it could be applied to a large class of chaotic dynamical systems.

\section{Taylor series method for the Lorenz system}
\label{sec:3}
We consider as a model problem the classical Lorenz system \cite{Lorenz}:
\begin{equation}
\begin{aligned}
\frac{dx}{dt} &= - \sigma x + \sigma  y\\
\frac{dy}{dt} &= Rx - y -xz\\
\frac{dz}{dt} &= xy - bz,
\end{aligned}
\end{equation}
where $R=28$, $\sigma=10$, $b=8/3$  are the standard Salztman's parameter values.
For these parameters the system is chaotic.
The N-th order Taylor series method \cite{Jorba}, \cite{Barrio} for (1) with step size $\tau$ is:
\begin{equation}
\begin{aligned}
x_{n+1} &= x_{n} + \sum_{i=1}^{N} \alpha_i \tau^i,\\
y_{n+1} &= y_{n} + \sum_{i=1}^{N} \beta_i \tau^i,\\
z_{n+1} &= z_{n} + \sum_{i=1}^{N} \gamma_i \tau^i,
\end{aligned}
\end{equation}
where
$$
\begin{aligned}
\alpha_i &= \frac{1}{i!}\frac{d^i x(t_n)}{{dt}^i},\\
\beta_i &= \frac{1}{i!}\frac{d^i y(t_n)}{{dt}^i}, \\
\gamma_i &= \frac{1}{i!}\frac{d^i z(t_n)}{{dt}^i}
\end{aligned}
$$
are the i-th Taylor coefficients (the so called normalized derivatives).
They are calculated as follows. From equation (1) we have
$$
\begin{aligned}
\alpha_1 &= -\sigma\alpha_0 + \sigma \beta_0,\\
\beta_1 &= R \alpha_0 - \beta_0 - \alpha_0 \gamma_0,\\
\gamma_1 &= \alpha_0 \beta_0 -b \gamma_0,
\end{aligned}
$$
where $$\alpha_0 = x_n, \hspace{0.5cm} \beta_0 = y_n, \hspace{0.5cm} \gamma_0 = z_n. $$

By applying Leibniz rule for the derivatives of the product of two functions,
we obtain the following recursive procedure for calculating  $\alpha_i, \beta_i, \gamma_i$  for $i=0,..., N-1$.

\begin{equation}
\begin{aligned}
\alpha_{i+1}  &= \frac{1}{i+1} (-\sigma \alpha_i +\sigma \beta_i),\\
\beta_{i+1}  &= \frac{1}{i+1} (R\alpha_i -\beta_i -\sum_{k=0}^{i}\alpha_{i-k}\gamma_k),\\
\gamma_{i+1}  &= \frac{1}{i+1} (\sum_{k=0}^{i}\alpha_{i-k}\beta_k -b\gamma_i).
\end{aligned}
\end{equation}

Note that we don't need any analytical expressions for the derivatives of $x(t)$, $y(t)$, $z(t)$
and we don't have any. We only need the values of the derivatives in point $t_n$.
Let's store the Taylor coefficients in the arrays \textbf{x},\textbf{y},\textbf{z} with lengths N+1. The values of $\alpha_i$ are stored in \textbf{x[i]},
those of $\beta_i$ in \textbf{y[i]} and those of $\gamma_i$ in \textbf{z[i]}. Then the pseudocode for (3) is given in Fig.1.
It is obvious  that we need $O(N^2)$ floating point operations for calculating the coefficients. The subsequent evaluation of Taylor series with Horner's rule needs only $O(N)$ operations and hence is negligible. To calculate the i+1-th coefficient in Taylor series we need all previous coefficients from 0 to i.
What we can do in parallel is the calculation of sums \textbf{s1} and \textbf{s2}, i.e. we can parallelize the loops for index \textbf{k}
(the code in frame in Fig. 1). This is a classical parallel reduction.

Actually, the explained above algorithm for calculating the coefficient of Taylor series,
is called automatic differentiation, or sometimes called algorithmic
differentiation (see \cite{Moore}).
Generally speaking, automatic differentiation is a recursive procedure for computing the derivatives of certain functions
at a given point (not  evaluating analytical formulas for the derivatives).
The certain functions are those that can be obtained by sum, product,
quotient, and composition of some elementary functions. It is important that in all cases of dynamical systems whose right-hand side is
automatically differentiable,  sums like those in (3) are obtained. Thus, the approach for parallelization of our model problem
can be applied straightforwardly to a large class of dynamical systems.

\begin{figure}
\begin{center}

\includegraphics[scale=.6]{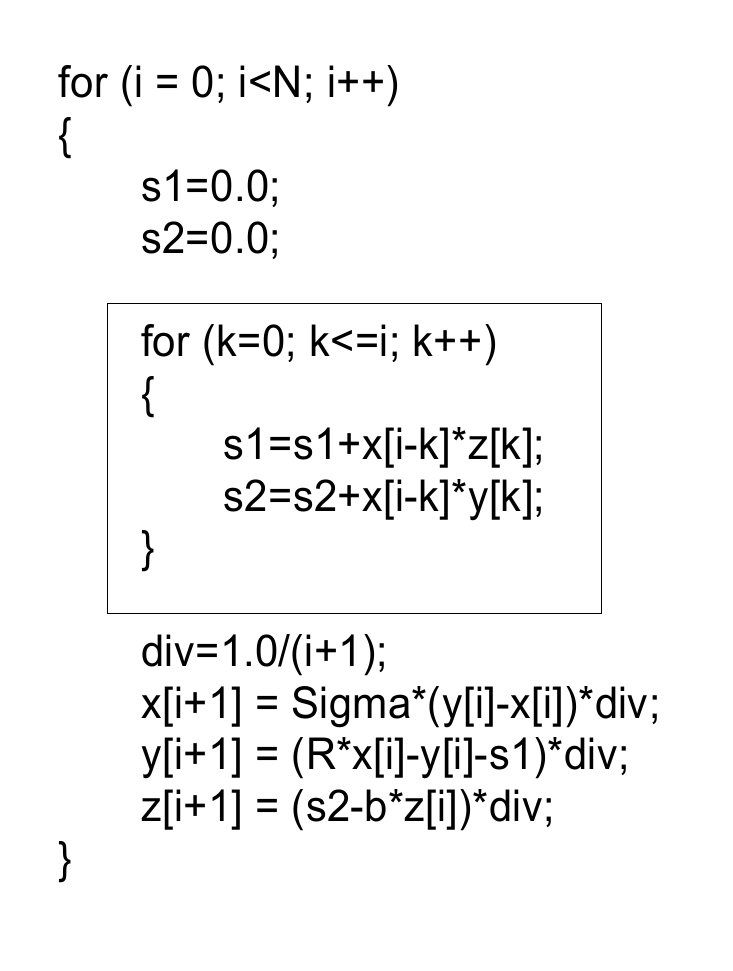}
\caption{Pseudocode of the recursive procedure (3).}
\label{fig:1}
\end{center}
\end{figure}

\section{OpenMP parallel technology}
Modern supercomputer clusters have nodes consisting of multicore
processors. Our goal is an effective and moderate (using only one node)  parallelism.
Parallel computing on one node  performs efficiently when a shared memory, multithreaded
environment, such as OpenMP, is used \cite{open}. OpenMP is an application programming interface (API) consisting
of a set of compiler directives (pragmas in C/C++) and a library of support functions.

The standard view of parallelism in OpenMP is the fork-join model.
When the program starts, only a single thread, called the master thread, is active.
The master thread performs the sequential parts of the algorithm.
At those points where parallel operations are required, the master thread forks (creates additional threads).
The master and the additional threads work concurrently trough the parallel region (some block of code).
At the end of the parallel region, created threads are suspended, and the flow of control returns to the single master thread,
this is the join point.

A nice feature of OpenMP is that it supports incremental parallelism,
i.e. we transform the sequential program into a parallel program step by step, one block of code at a time.
If the main part of computational work is done only in one block, as in our case, adding one or two lines (pragmas)
in the code and eventual minor transformations, might be all we need for excellent speedup results.

\begin{figure}
\begin{center}
\includegraphics[scale=.6]{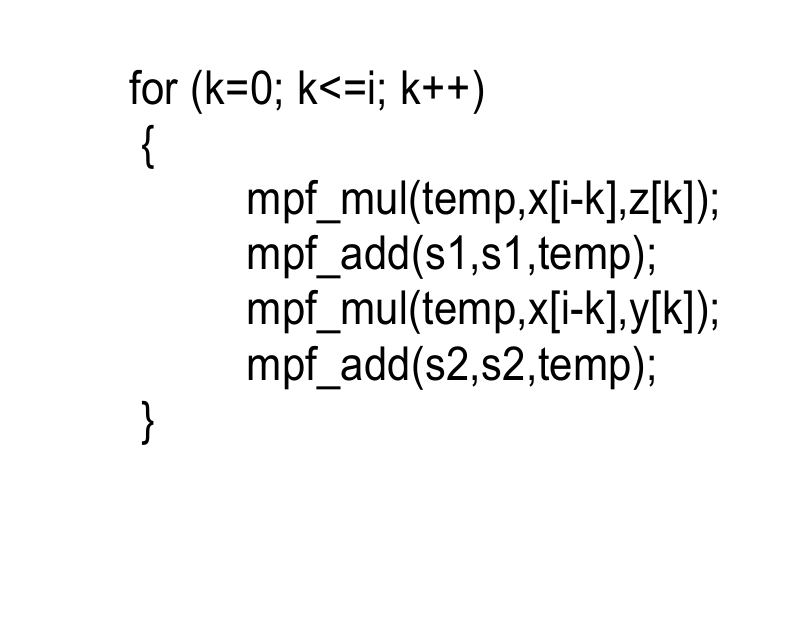}
\caption{The serial block of code that we have to parallelize, written in terms of GMP.}
\label{fig:2}
\end{center}
\end{figure}

\section{OpenMP parallel reduction}
We write our parallel program in C programming language and use the GMP library (The GNU Multiple Precision
Arithmetic Library) \cite{gnu}. An important feature of the GMP library is its thread-safety \cite{open}, otherwise we can't use it with OpenMP.
The serial block of code that we have to parallelize, written in terms of GMP, is given in Fig.2. We use a temporary variable \textbf{temp}
to store the intermediate results of the multiplications before additions to the sums \textbf{s1} and \textbf{s2}.

To parallelize this code, we have to do a parallel reduction. OpenMP has a build-in reduction clause, however we can't use it, because we use user-defined types for multiple precisions number and user-defined operations. Thus, we have to do the reduction manually. Because multiple precision
numbers are allocated in the heap, they can not be privatized \cite{open}, thus we have to make containers for the partial sum for every thread
and these containers will be shared. We store the containers in an array of multiple precision numbers \textbf{sum} with length \textbf{2Numt},
where \textbf{Numt} is the number of threads. Thread  with number \textbf{tid} stores its portion of \textbf{s1} in \textbf{sum[2*tid]} and its
portion of \textbf{s2} in \textbf{sum[2*tid+1]}. Now we have an array of temporary variables \textbf{tempv} with length \textbf{Numt}.

The parallelized version of the code from Fig.2 is given in Fig.3.
The directive \textbf{\#pragma omp parallel private(k,tid)} creates a parallel region (additional threads are awaken)
and the integer variables \textbf{k} and \textbf{tid} are defined as private in the parallel region (each tread has its own copy in its own stack).
Then every thread gets its \textbf{ID} and stores it in \textbf{tid} by using the library function \textbf{omp\_get\_thread\_num()}.
The second directive \textbf{\#pragma omp for} shares the work for the loop
between threads, i.e. every thread works on its own portion of the range of indexes [0,...,i]. This range is divided in \textbf{Numt} chunks, equal in size as much as possible. The first right bracket closes the \textbf{for} loop section and acts as an implicit barrier for synchronization of  the threads. The second right bracket closes the entire parallel region (the join point). Out of the parallel region the master thread unloads all containers, belonging to each thread, over \textbf{s1}, \textbf{s2}.

\begin{figure}
\begin{center}
\includegraphics[scale=.6]{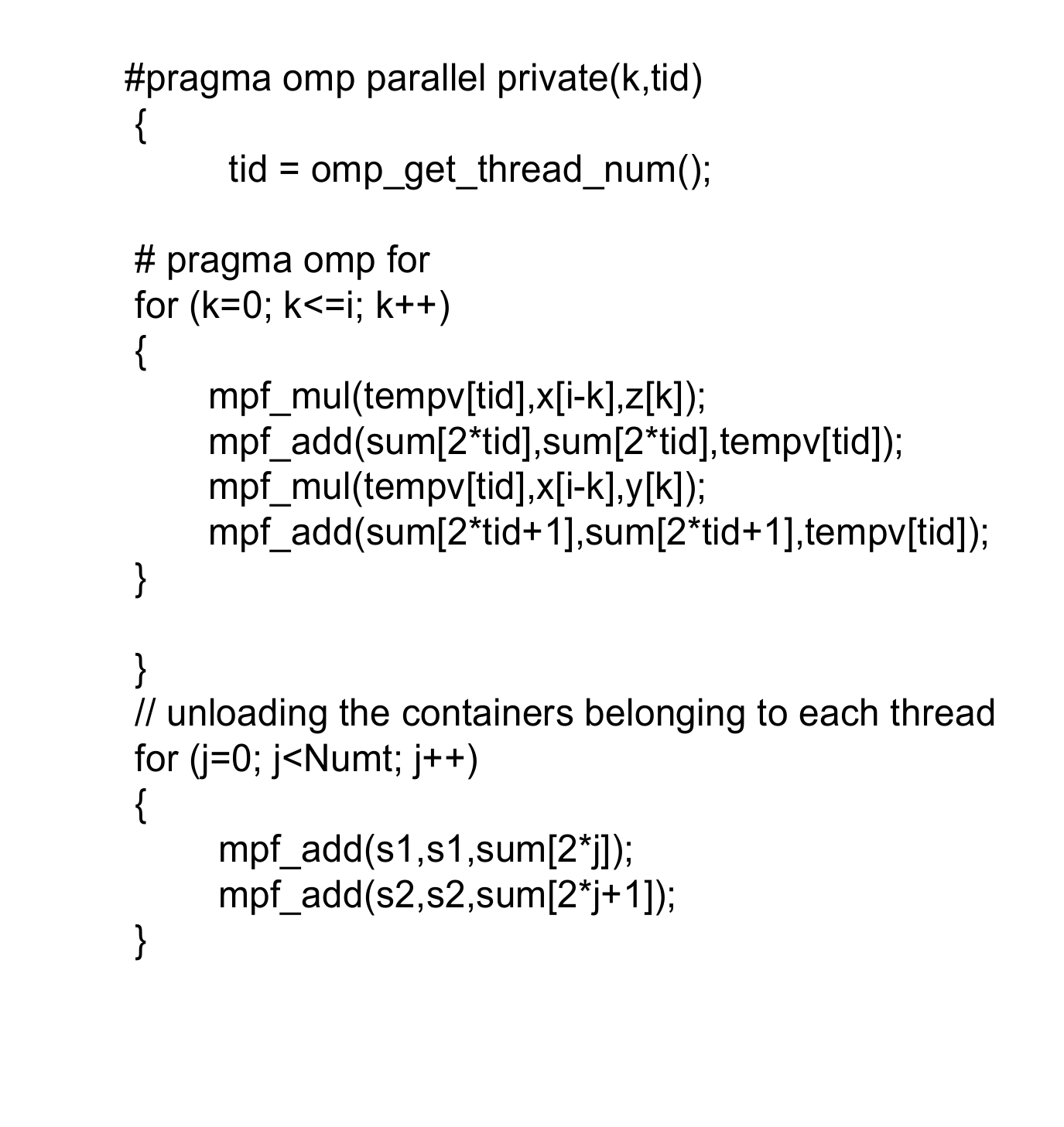}
\caption{The parallelized block of the code.}
\label{fig:3}
\end{center}
\end{figure}

\section{Performance results}
All computations are performed  on the  HybriLIT Heterogeneous Platform at the Laboratory of IT of JINR, Dubna, Russia \cite{hlt}, where the GMP library (version 6.1.2) is installed. We use one computational CPU-node consisting of 2 x Intel(R) Xeon(R) Processor E5-2695v3 (28 cores, 2.3 GHz).

As a benchmark we use the results for the Lorentz equations with initial conditions and step size taken from \cite{par2}, namely $x(0)=-15.8$, $y(0)=-17.48$,
$z(0)=35.64$, $\tau=0.01$. First, we compare our performance with those in Table 1 in \cite{par2}. The results in this table are for a problem with $t\in [0,1200]$ that used order of Taylor series $N=400$ and precision with $K=800$ decimal digits.
The  number of CPU-cores used in \cite{par2} is between 1 and 50. We did not expect too much parallel efficiency on all 28 cores for this relatively small
problem. Our serial time is $\sim$ 9.0 hours and it is similar to that in \cite{par2} $\sim$ 9.8 hours. However, our program shows better parallel efficiency on this small problem. We obtain about 44 \% parallel efficiency on 28 cores comparing with the same efficiency on 20 cores in \cite{par2}.  Our time on 28 cores is $\sim0.73$ hours which is even better than the time $\sim0.83$ hours on 50 cores in \cite{par2}.

To make a simulation in the rather large interval [0,5000], we use order of method $N=2000$ and precision of 8000 bits ($K\sim$ 2412 decimal digits), according to $T_c - K$ and $T_c - N$ diagrams in \cite{Liao1}. We test the performance scalability and efficiency inside the CPU node with a short length  simulation test (5 integration steps).
The results in Table 1 show a very good performance scalability and efficiency about 75\% on all 28 cores.
In a foreseeable time ($\sim$ 9 days and 14 hours) on 28 cores we integrated the equations in the time interval [0,5000] and repeated the benchmark table S1 from paper \cite{par2}.

\begin{table}
\caption{OpenMP parallelization of 2000-th order Taylor method with precision $\sim$ 2412 decimal digits on the example of the Lorenz equations.
 The computations are made on one CPU-node consisting of 2 x Intel(R) Xeon(R) Processor E5-2695 v3 (28 cores, 2.3 GHz).
 The measured time is for a short length  simulation test (5 integration steps).}
\label{tab:1}

\begin{tabular}{p{2.7cm}p{2.9cm}p{2.9cm}p{2.8cm}}
\hline\noalign{\smallskip}
Number of cores & Time (seconds) & Speedup  & Parallel efficiency (\%)  \\
\noalign{\smallskip}\svhline\noalign{\smallskip}
1 (serial) & 174.05 & 1.00 & 100.0\\
1 (parallel) & 174.57 & 0.997 & 99.7\\
4 & 47.89 & 3.63 & 90.9\\
8 & 25.16 &  6.92 & 86.5\\
14 & 15.32 &  11.4 & 81.1\\
28 & 8.28 &  \textbf{21.0} & 75.1\\
\noalign{\smallskip}\hline\noalign{\smallskip}
\end{tabular}
\end{table}

\section{Conclusions}
OpenMP parallelization of multiple precision Taylor series method for the Lorenz system is realized.
A very good parallel performance scalability and parallel efficiency inside one computation node of a CPU-cluster is observed.
Our approach for parallelization can be simply applied  to a large class of chaotic dynamical systems
in order to obtain long-term mathematically reliable solutions.

\begin{acknowledgement}
We thank the Laboratory of Information Technologies of JINR, Dubna, Russia for the opportunity to use the computational resources of the HybriLIT Heterogeneous Platform.
The work is financially supported by a grant of the Plenipotentiary Representative of the Republic of Bulgaria at the JINR.
\end{acknowledgement}

\end{document}